\documentclass[%
superscriptaddress,
 preprint,
amsmath,amssymb,
aps,
longbibliography,
]{revtex4-1}

\usepackage{array}
\usepackage{graphicx}
\usepackage{float}

\begin{document}

\title{Diffusion-limited settling of highly porous particles in density-stratified fluids}

\author{Robert Hunt}
\email{robert\_hunt@brown.edu, daniel\_harris3@brown.edu}
\affiliation{Center for Fluid Mechanics, School of Engineering, Brown University}

\author{Roberto Camassa}
\affiliation{Department of Mathematics, The University of North Carolina at Chapel Hill}
\author{Richard M. McLaughlin}
\affiliation{Department of Mathematics, The University of North Carolina at Chapel Hill}
\author{Daniel M. Harris}
\email{robert\_hunt@brown.edu, daniel\_harris3@brown.edu}
\affiliation{Center for Fluid Mechanics, School of Engineering, Brown University}

\date{September 3, 2024}

\begin{abstract}
The vertical transport of solid material in a stratified medium is fundamental to a number of environmental applications, with implications for the carbon cycle and nutrient transport in marine ecosystems. In this work, we study the diffusion-limited settling of highly porous particles in a density-stratified fluid through a combination of experiment, analysis, and numerical simulation. By delineating and appealing to the diffusion-limited regime wherein buoyancy effects due to mass adaptation dominate hydrodynamic drag, we derive a simple expression for the steady settling velocity of a sphere as a function of the density, size, and diffusivity of the solid, as well as the density gradient of the background fluid. In this regime, smaller particles settle faster, in contrast with most conventional hydrodynamic drag mechanisms. Furthermore, we outline a general mathematical framework for computing the steady settling speed of a body of arbitrary shape in this regime and compute exact results for the case of general ellipsoids. Using hydrogels as a highly porous model system, we validate the predictions with laboratory experiments in linear stratification for a wide range of parameters. Lastly, we show how the predictions can be applied to arbitrary slowly varying background density profiles and demonstrate how a measured particle position over time can be used to reconstruct the background density profile. 

\end{abstract}

\maketitle

\section*{Introduction}
The settling of solid particles in a fluid is one of the most fundamental problems in fluid dynamics, with applications spanning many scales and scientific fields. One important application of recent interest involves particles settling in the ocean: marine snow, which refers to the transport of organic material introduced near the surface, plays an essential role in the carbon cycle and in the ocean ecosystem. Further, human generated waste and microplastics are found in all corners of the Earth, yet the mechanisms that affect their dispersion are not well understood \cite{sutherland2023}.

Understanding the distribution and transport of these materials is essential for predicting and controlling carbon sequestration and microplastic dispersion. Many organic materials are well represented as porous particles, and particles which float at the ocean surface are generally known to develop coatings of porous and organic material due to biofouling before sinking \cite{fazey2016,semcesen2021,liu2022}. Smaller particles and particles with higher surface area to volume ratios are speculated to leave the surface sooner and sink faster, but this has not been studied in depth. The distribution of these porous particles is also closely linked with ocean ecology and has been shown to be associated with increased biological activity \cite{smith1992,PRAIRIE2015,Prarie2017}.

The seminal work by G.G. Stokes \cite{stokes1851} provided an expression for the drag on a sphere in a viscous fluid, which, when balanced with gravity, yields the celebrated Stokes settling law, $U_{\mu} = \frac{2 \rho_s g R^{2}}{9 \mu}$, where $U_{\mu}$ is the particle settling velocity, $\rho_s$ is the relative density difference of the particle to the uniform background fluid, $g$ is the gravitational acceleration, $R$ is the particle radius, and $\mu$ is the dynamic viscosity of the fluid. This work was expanded experimentally through the 1900s across a large range of Reynolds numbers and has continued to find applications across scales and disciplines. This characterization was extended by \cite{zvirin1975, abaid2004, yick2009, camassa2009, camassa2010, mehaddi2018,Zhang2019,Camassa2022} to include effects of ambient density stratification, a topic which has remained an active area of research and was summarized in a recent review article \cite{more2023}. Stratification is known to strongly influence individual particle settling behavior as well as drive particle aggregation due to diffusion-induced flow \cite{camassa2019}, with the predicted behaviors depending on porosity. In addition, the total particle mass can change in time as it moves through the stratified environment, dramatically affecting the particle's steady settling behavior in certain regimes. However, there is very limited work on porous particles settling through stratification in cases where the background stratification 
changes over length scales much larger than the size of the settling particles despite this representing the regime most relevant to the aforementioned environmental applications. Before moving to the specific focus of the present work, we briefly review some of the most relevant literature in what follows.

In early field work, Alldredge and Gottschalk \cite{alldredge1987} measured the in-situ settling of marine snow particles off of the Southern coast of California, with attempts to measure excess density, porosity, and volume, and noted that the settling did not obey Stokes' law. They also noted the shapes of the settling aggregates were often non-spherical, with a tendency to be axisymmetric and elongated along the settling direction. In similar regions,  MacIntyre et al. \cite{macintyre1995} documented large accumulations of marine snow at pycnoclines and proposed diffusive mass exchange into the highly porous flocs as a potential mechanism for such increased retention.

In the lab, Li et al. \cite{li2003} considered the effect of mass-exchange on the settling of biological aggregates through a sharp, two-layer stratification. They noted that, in the bulk of each layer, the aggregated particles were well-approximated by Stokes' settling law. They also investigated the residence time of particles at the interface between the two layers and compared that with a proposed theoretical model for the retention time. Their model predicted the retention time to scale as the particle radius squared, although they had mixed agreement with experiments. 

Several years later, Kindler et al. \cite{kindler2010} performed similar settling experiments of spherical porous hydrogel particles in a two-layer fluid and confirmed the previously posited quadratic scaling of retention time with respect to particle radius. They also proposed a quasi-static model for the particle position versus time, accounting for particle inertia, mass adaptation, and form drag, which correctly predicted the strong deceleration and shape of the particle trajectory but somewhat overpredicted the settling velocity and mass adaptation away from the center of the pycnocline. In the discussion, a scaling argument for the {\it steady} settling velocity of a porous sphere in the diffusion-limited regime was derived, by suggesting that the density rate of change of the particle must match that of the background fluid. This scaling law was not directly tested.

In closely related work, Camassa et al. \cite{camassa2013} also performed experiments of porous particles settling through a two-layer fluid and showed good agreement with a first-principles model that includes the effects of mass exchange and Stokes drag. They accounted for the effect of the entrained lighter fluid, initially investigated experimentally as a competing effect by Prarie et al. \cite{prarie2013}, by introducing a single adjustable parameter that represents a thin uniform shell surrounding the particle. They noted that for small particles, the settling time is dominated by viscous drag (with settling time scaling as $R^2$), whereas for large particles the settling time is dominated by mass diffusion into the sphere (with settling time scaling as $R^{-2}$). Additional numerical simulations of this problem were completed by Panah et al. \cite{panah2017}, who proposed various empirical laws for the settling time in a two-layer system, with additional characterization on the influence of the thickness of the transition layer separating the two fluids.

While the prior modeling and laboratory studies discussed above predominantly focused on spheres in relatively sharply stratified two-layer systems, there has only been very limited attention to the case of linear stratification, or more generally, in cases where the stratification changes over length scales that are significantly larger than the particle. Very recently, Ahmerkamp et al. \cite{ahmerkamp2022} considered the settling of spherical porous particles in linear stratification both numerically and experimentally, motivating their exploration by the large length scales of stratification typically found in environmental settings. In their simulations, they considered a relatively general set of parameters and noted that the expected settling velocities, as one might estimate from hydrodynamic drag alone, can be notably reduced by the effect of mass exchange between the background, boundary layer, and particle interior. Various empirical correlations for the drag force were determined by fitting to their numerical results. However, there is only limited comparison to experimental results and with mixed success, and their study focuses strictly on spherical particles. Nevertheless, this work clearly highlights the richness and complexity of the general problem due to the highly coupled multi-physics involved and the importance of considering mass exchange in estimating steady settling velocities.

In the present work, we study the steady settling of highly porous particles in linear stratification. We consider these particles as solids that allow for diffusion of a solute through their interior but are inpenetrable to fluid flow. This focus is similarly motivated by marine systems where aggregates are generally of very high porosity ($\gg$ 95\%) \cite{ahmerkamp2022,macintyre1995,alldredge1988} and density gradients may vary on the order of meters to kilometers. However, we focus herein on the diffusion-limited regime, which we demonstrate can be delineated by an appropriately defined non-dimensional Rayleigh number. In this regime, we are able to derive an exact formula for the particle settling speed from first principles as well as generalize the result to bodies of arbitrary geometry. Extensions and applications to nonlinear stratifications are also explored. 

The theoretical predictions are validated by controlled laboratory experiments that directly measure the steady settling velocities of agar particles immersed in a linear stratification of salt (sodium chloride). Agar is a highly porous material (composed of mostly interstitial water) that permits diffusive transport of salt into its bulk but is essentially impermeable to fluid flow. The agar particles are cast in 3D-printed molds, placed gently in a stably stratified tank of fluid, and visualized from the side as they slowly fall. Further details on the experimental methods are provided in the Methods section.

\section*{Results}

\begin{figure*}
\centering
\includegraphics[width=\textwidth]{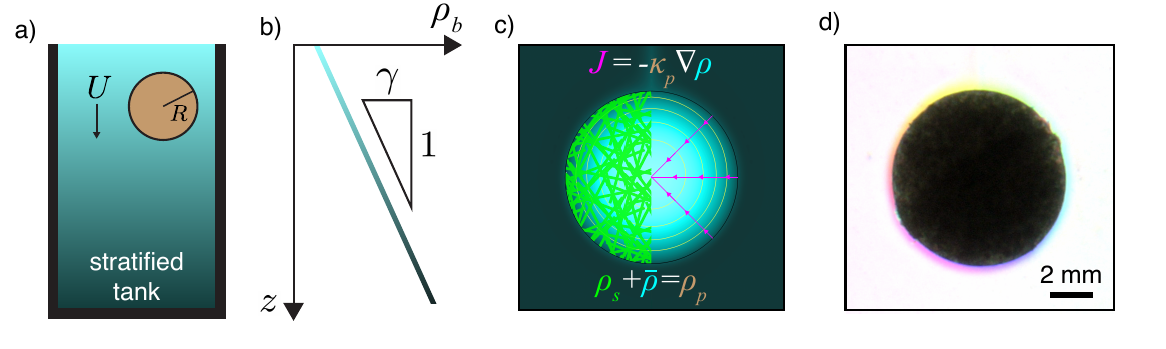}
\caption{a) Depiction of a highly porous sphere (brown) of radius $R$, density $\rho_p$, and diffusivity $\kappa_p$ settling in a density-stratified tank with settling velocity $U$. b) Stable linear density stratification profile with vertical density gradient 
$\gamma$. c) Steady solute concentration field difference from linear background obtained from simulations. The solid density $\rho_s$ due to the solid agar matrix, assumed to be homogeneous, is represented abstractly in green. The solute density field $\rho$ is represented in blue, with lighter color corresponding to a lower solute concentration and corresponding density as in panels a) and b). The volume average of this internal solute field yields the total solute density $\bar{\rho}$. Isolines of solute concentration are pictured in yellow, and the solute flux $J$ is represented by streamlines in magenta. The solute evolves diffusively inside the particle with an effective diffusivity $\kappa_p$, and the sum of $\bar{\rho}$ and $\rho_s$ gives the total particle density $\rho_p$.  d) Image of actual particle from experiments with visualization of exterior density field gradient using a color-based Schlieren method. The color saturation indicates the magnitude of image distortion due to variations in the index of refraction, highlighting the relatively thin convective layer of lower density surrounding the particle exterior.}
\label{fig:Setup}
\end{figure*}

\subsection*{Model}
Consider a spherical particle of radius $R$ that settles vertically in a stable background stratification profile $\rho_b(z)$ with constant density gradient $\gamma=\frac{d\rho_b}{dz}$, as depicted in Figure \ref{fig:Setup}(a,b). The particle is assumed impermeable to fluid flow but diffusively permeable to the stratifying agent such that its mean density can change in time. The equations describing the fully coupled fluid-solute-particle system have been previously outlined in prior work \cite{ahmerkamp2022,panah2017} but will be described briefly here. The interior of the particle is described by a diffusion equation for the solute, whereas the exterior fluid domain is modeled by the Navier-Stokes equations and an advection-diffusion equation for the stratifying solute. The interior and exterior domains are coupled by enforcing continuity of stress, velocity, solute concentration, and solute flux at the particle boundary. For the general case, the fully coupled system requires numerical solution. However, we seek to simplify the full system to a reduced model by appealing to specific parametric regimes, following Camassa et al. \cite{camassa2013}. In this formulation, the boundary conditions associated with the exterior problem are greatly simplified by assuming the fluid stresses to be approximated by classical Stokes flow and buoyancy due to the undisturbed background stratification, with the exterior solute concentration approximated by the value of the background concentration profile evaluated at the center height of the particle (the so-called ``heat bath'' approximation). Under these assumptions, the problem can be simplified to an equation for the position of the particle's center $Z(t)$, representing a force balance between Stokes drag and buoyancy, coupled with a diffusion equation for the evolution of the solute in the particle interior whose boundary condition is given by the background density evaluated at the height corresponding to the particle center. For the case of a sphere of radius $R$, this corresponds to
\begin{equation}
\label{eq:forceBalance}
    6 \pi \mu R \frac{d}{dt} Z(t) =  g V \left[\rho_{p}(t)-\rho_{b}(z=Z(t))\right]
\end{equation}
where $\mu$ is the dynamic viscosity of the fluid, $g$ is the acceleration due to gravity, and $V=\frac{4}{3}\pi R^3$ is the particle volume. The density $\rho_{b}(z=Z(t))$ is the density of the stratified background fluid at the height of the particle center, whereas $\rho_{p}(t)$ is the average total density of the particle itself. For a fixed particle density, this represents the Stokes settling of an impermeable solid particle in a background linear stratification and has been the subject of prior analysis \cite{zvirin1975}. In this work, the particle is considered to be a very low volume fraction solid material (i.e. highly porous) that allows for diffusion of the stratifying solute with an effective diffusivity $\kappa_p$ throughout its bulk. The extent of fluid flow through the particle itself can be characterized by the Darcy number, defined as $Da = \frac{K}{R^{2}}$, where $K$ represents the particle permeability. Here, we appeal to the limit of small $K$ and $Da$ and neglect fluid flow through the solid medium as in prior studies \cite{kindler2010,panah2017}. Thus the total density of the particle can be considered as
\begin{equation}
\label{eq:particleDensity}
    \rho_{p}(t) = \rho_{s} + \bar{\rho}(t)
\end{equation}
where $\rho_{s}$ is a fixed material parameter representing the contribution to the total density from the solid material and
\begin{equation}
\label{eq:particleDensityAverage}
    \bar{\rho}(t) = \frac{1}{V}\int_{\Omega} \rho(\textbf{r},t)
\end{equation}
is the average density of the solute field $\rho(\textbf{r},t)$ in the particle interior, given by the solution of the diffusion equation
\begin{equation}
\label{eq:diffusion}
    \frac{d}{d t}\rho(\textbf{r},t) = \kappa_p \nabla^2 \rho(\textbf{r},t)
\end{equation}
with boundary condition
\begin{equation}
\label{eq:BC}
    \rho(\textbf{r},t)|_{\delta\Omega} = \rho_{b}(Z(t))
\end{equation}
where $\delta \Omega$ represents the particle boundary.

Furthermore, we will consider the case of a linear background density stratification
\begin{equation}
\label{eq:bgdensity}
    \rho_{b}(z) = \gamma z
\end{equation}
where the parameter $\gamma>0$ defines the constant density gradient, representing  a stable density configuration with the density increasing with depth. Note that $\rho_{b}(z)$ represents the excess fluid density associated with the presence of the stratifying agent, such that in a homogeneous fluid $\rho_{b}(z)=0$. Equation \ref{eq:forceBalance} can be recast in non-dimensional form as
\begin{equation}
\label{eq:camassa1}
    \frac{9}{2Ra} \frac{d}{dt^*} Z^*(t) =  \rho^*_{p}(t) - Z^*(t)
\end{equation}
where $Z^*(t) = Z(t)/R$, $t^* = t \kappa_p / R^2$, $\rho^*_{p} = \rho_{p}/\gamma R$, and $Ra=\frac{g \gamma R^4}{\kappa_p \mu}$. The solutal Rayleigh number $Ra$ represents a balance of buoyant to viscous forces. We can find steady solutions to this reduced system by assuming a constant speed $U$. In the frame of the sphere $\Tilde{z} = z - U t$, $\rho_{b}(\Tilde{z}) = (\Tilde{z} + U t) \gamma $, and $\Tilde{\rho} = \gamma U t + f(\mathbf{r})$, so that $f(\mathbf{r})$ satisfies the Poisson equation
\begin{equation}
\label{eq:diffusionsteady}
    \gamma U = \kappa_p \nabla^2 f(\mathbf{r}),
\end{equation}
with homogeneous boundary condition
\begin{equation}
\label{eq:diffusionsteady2}
    f(\mathbf{r})|_{\delta \Omega} = 0.
\end{equation}
For the case of the sphere, the solution is $f(\mathbf{r}) = \frac{\gamma U}{6 \kappa_p}(r^{2}-R^{2})$. By averaging over the volume of the sphere, we can find an exact expression for the mean excess density due to the solute, $\bar{\rho}(t) = \gamma U t - \frac{ \gamma U R^2 }{ 15 \kappa_p}$. The linearly growing part of this expression is balanced with the background density, and the constant term represents a stable contribution due to the lower mean solute concentration in the particle interior, which is balanced by the extra solid density $\rho_s$ and fluid drag.

Then returning to Equation \ref{eq:camassa1}, we find
\begin{equation}
\label{eq:Ufull}
U = \frac{15 \kappa_p \rho_{s}}{\gamma R^{2}}\left(1 + \frac{2}{135Ra}\right)^{-1}.
\end{equation}
This solution can also be represented as a parallel sum between settling velocities at the two limiting behaviors of the system: 
\begin{equation}
 \label{eq:Usum}
 U = \left(\frac{1}{U_{\gamma}}+\frac{1}{U_{\mu}}\right)^{-1}
 \end{equation}
where $U_{\gamma} = \frac{15 \kappa_p \rho_{s}}{\gamma R^{2}}$ is the diffusion-limited settling velocity for $Ra \gg 135/2$ and $U_{\mu} = \frac{2 g \rho_{s} R^2}{9 \mu}$ is the Stokes settling velocity for $Ra \ll 135/2$. For more details regarding the model and derivation, see the Supplementary Information. In this work, we specifically focus on the diffusion-limited regime corresponding to large $Ra$, where the fluid drag can be effectively ignored and the settling dynamics are governed by mass exchange. This limit ultimately leads to a simple expression for the steady settling speed of a spherical particle in the diffusion-limited regime:
\begin{equation}
\label{eq:ugamma}
U = \frac{15 \kappa_p \rho_{s}}{\gamma R^{2}}.
\end{equation}
This equation suggests that larger particles will settle slower than otherwise equivalent small particles, in direct contrast to the Stokes regime. While a similar scaling was proposed by Kindler \cite{kindler2010}, to the best of our knowledge this is the first analytical expression derived from first principles for the diffusion-limited settling velocity of a particle falling in linear stratification.

The solutal Rayleigh number $Ra$ can also be interpreted as a ratio of timescales $t_{\gamma}/t_{\mu}$, where $t_{\gamma} = R^2/\kappa_p$ represents the characteristic diffusion time and $t_{\mu} = \frac{\rho_{s}}{\gamma U_{\mu}}$ represents the time for a Stokes particle to settle a characteristic distance $\rho_{s}/\gamma$. This distance can be interpreted as the characteristic vertical distance the particle must fall in order for the excess background density to become comparable to the excess solid density of the particle. In the diffusion-limited regime ($Ra \gg 135/2$), these timescales are well separated, with the initial transient dynamics depending both on the characteristic diffusion time of the particle $t_{\gamma} = R^2/\kappa$ and the initial position of the particle. At short times, the particle density will not change appreciably (as diffusion has not had time to act) and at first order will exponentially approach its equilibrium height (assuming Stokes drag or asymptotic corrections thereof, as discussed in \cite{zvirin1975}). Transient dynamics due to the initial condition of the solute will then decay on a timescale according to $t_{\gamma}$ as the particle then begins its steady diffusion-limited descent (for details see the Supplementary Information).

Our initial assumption of the external fluid stresses arising principally from Stokes drag neglects effects of fluid inertia and the possibility of additional buoyancy due the solute evolution in the background fluid that can take the form of a density boundary layer. Although these effects are not negligible in all cases, we expect the diffusion-limited behavior characterized in this work to apply to any scenario where hydrodynamic drag can be neglected and the density boundary layer surrounding the particle is small relative to the particle size. By restricting focus to this purely diffusion-limited regime, we can in fact readily generalize Equation \ref{eq:ugamma} to any arbitrary geometry via
\begin{equation}
\label{eq:exactsol}
    U = \frac{15 \kappa_p \rho_{s}}{\gamma R_{e}^{2}} 
\end{equation}
where $R_{e}$ is the effective radius of the particle. For a general solid volume, $R_{e}=\sqrt{-15\bar{\phi}}$, where $\bar{\phi}$ is the volume average of the solution to the Poisson problem $\nabla^2 \phi=1$, $\phi|_{\delta\Omega}=0$ (for more details see the Supplementary Information). For a sphere, $R_{e}$ is simply the sphere radius $R$, consistent with Equation \ref{eq:ugamma}. While the problem can be readily solved numerically for an arbitrary three-dimensional geometry, certain geometries admit analytical solutions. For instance, in the case of a general ellipsoid, one can find a steady solution for the density field in the interior of the particle as 
\begin{equation}
\label{eq:ellipsoiddensity}
    f(\mathbf{r}) = \frac{\gamma  U}{2
   \kappa_p}\left(\frac{1}{a^2}+\frac{1}{b^2}+\frac{1}{c^2}\right)^{-1}\left(\frac{x^2}{a^2}+\frac{y^2}{b^2}+\frac{z^2}{c^2}-1\right),
\end{equation}
which corresponds to an effective radius of $R_{e}$ of
\begin{equation}
\label{eq:ellipsoidshapefactor}
    R_{e} = \sqrt{3\left(\frac{1}{a^{2}}+\frac{1}{b^{2}}+\frac{1}{c^{2}}\right)^{-1}}
\end{equation}
where $a$, $b$, and $c$, are the semi-axis lengths of the ellipsoid. More details regarding the general ellipsoid and other geometries are provided in the Supplementary Information.

The predictions for the diffusion-limited settling velocities for both spheres and ellipsoids will be tested experimentally in what follows. Specific trials are compared to simulations of the fully-coupled system developed in COMSOL. For more details regarding experimental and simulation methods, see Methods.

\subsection*{Spheres}

\begin{figure*}
\centering
\includegraphics[width=.95\textwidth]{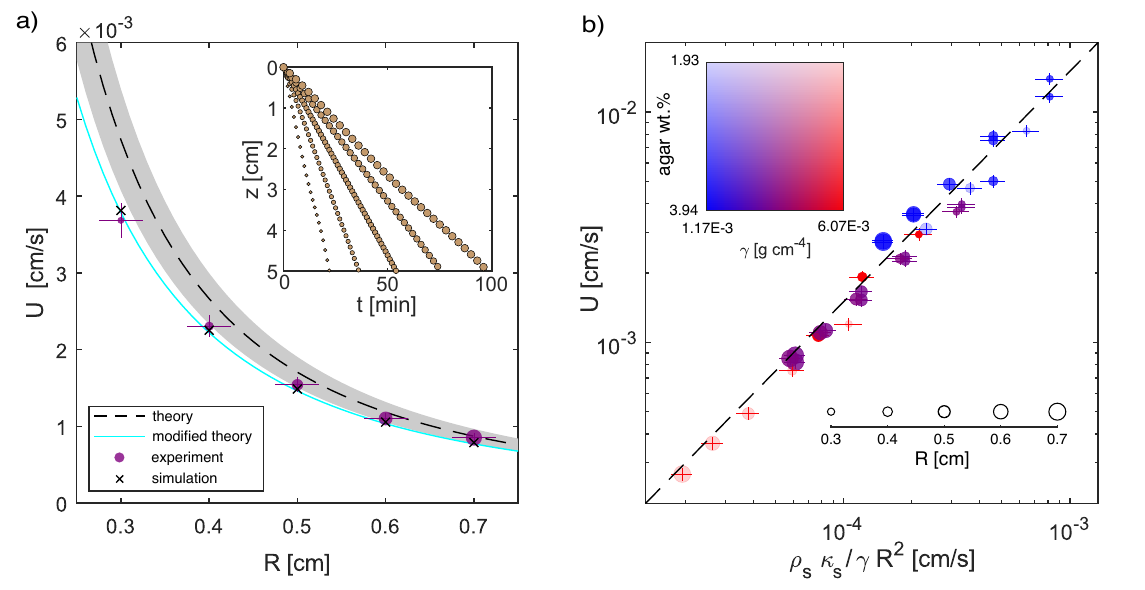}
\caption{a) Dimensional plot of agar sphere settling velocity $U$ versus radius $R$ for $R = 0.3 -0.7$ cm, agar wt.\% = 3.44, $\rho_s = 1.0\times10^{-2}$ g cm$^{-3}$, $\kappa_p = 1.0\times10^{5} $ cm$^2$ s$^{-1}$, and $\gamma = 3.7\times10^{-3}$ g cm$^{-4}$ ($Ra = 0.8 - 2.4 \times 10^{7}$). Inset shows corresponding particle positions versus time over the 5 cm region of measurement, with corresponding raw data included in Movie S1. The theoretical prediction (Equation \ref{eq:ugamma}) is represented by a black dashed line with propagated uncertainty associated with uncertainty in measured parameters shown in grey. The modified theory accounting for the depleted convective region of width $\alpha\delta$ is pictured in cyan. Predictions from numerical simulations are depicted by a black $\times$. Experimental results are purple, corresponding to their percentage of agar by weight and density gradient as characterized by the panel b) inset, with error bars representing uncertainty in $R$ and $U$. b) Non-dimensional collapse of experimentally measured sphere settling velocity $U$ versus a combination of measured parameters that scale with the theoretical prediction for settling velocity (Equation \ref{eq:ugamma}, represented by the black dashed line). Marker size indicates sphere radius, color indicates strength of gradient, and opacity represents the agar wt.\% as shown in the legends. Raw data included in SI dataset S1 \& S2.}
\label{fig:SpheresDimensional}
\end{figure*}

A typical trial is shown in Figure \ref{fig:SpheresDimensional}(a), with raw data included in Movie S1. Here, five agar spheres of varying radius are measured settling in a linearly stratified tank. Other than the particle radius, all other parameters are held fixed. One can clearly see that the settling speed depends inversely on the size, with the smallest sphere settling most rapidly. The measured velocities also agree reasonably well with the prediction of Equation \ref{eq:ugamma}, although are consistently overpredicted. The discrepancy between the analytical solution and experiment is relatively small and reduces as the particle size increases. Note that the shaded band surrounding theoretical prediction represents the propagation of uncertainty on the measured parameters that contribute to the prediction in Equation \ref{eq:ugamma}. The simulation results of the fully coupled system are in excellent quantitative agreement with experiment.

We suggest that the remaining discrepancy is due to the convective fluid boundary layer of reduced density fluid just outside of the solid particle \cite{camassa2013, ahmerkamp2022}, which we have neglected in the simplifications leading up to Equation \ref{eq:ugamma}. In our regime, this layer has the primary effect of furthering delaying the mass adaptation of the solid. To explore this difference, we roughly estimate the thickness $\delta$ of this convective layer using classical results for free convection from a vertically oriented heated plate \cite{ostrach1953}, known to have a boundary layer thickness $\delta$ that scales along the length of the plate $x$ as $\delta \sim  \left(\frac{4 \nu^2 x}{g \Delta \rho}\right)^{1/4}$. By assuming a characteristic length scale $x \sim R$ and characteristic density difference $\Delta \rho \sim \rho_s$, we find an approximate scaling for this boundary layer thickness in our problem as
\begin{equation}
\label{eq:boundarylayer}
    \delta \sim \left(\frac{4 \nu^2 R}{g \rho_{s}}\right)^{1/4}.
\end{equation}
In general, the solute can be transported by free and forced convection. The Richardson number represents the relative magnitudes of free to forced convection and is large for our experimental regime  ($Ri  =\frac{g \rho_s R}{U_{\gamma}^2} = O(10^{5})$), justifying the assumption of free convection in our estimate. Thus, we might introduce a `virtual' boundary in the simplified model, as in \cite{camassa2013}, by extending the effective solid boundary so that it includes the convective layer (i.e. $R\rightarrow R + \alpha \delta$). The prefactor $\alpha$ is unknown but presumably of $O(1)$. A best fit correction to our experimental velocity data for this trial yields a prefactor of $\alpha=0.60$. More general empirical relations on this buffering boundary layer are detailed in prior work \cite{ahmerkamp2022}. For all of our experimental trials $\delta/R < 0.21$, suggesting that the influence of the boundary layer on the diffusive transport of solute into the sphere is a secondary effect, and thus we choose to neglect it henceforth. Furthermore, its estimated influence is of similar order to our prediction uncertainty associated with the propagated uncertainty of the parameters.  This boundary layer is fully resolved in our COMSOL simulations and is also small relative to the sphere size in all cases.

In Figure \ref{fig:SpheresDimensional}(b), all experimental trials conducted for spheres in linear stratification are plotted and compared with the theoretical prediction of Equation \ref{eq:ugamma}. Good agreement is demonstrated across orders of settling velocity, for multiple values of sphere density, diffusivity, and background gradients. For all cases considered here, the observed settling velocity is within 27\% of the prediction for $U_{\gamma}$, with a mean error of 13\%. On average, the spheres move 5.9\% slower than predicted.

\subsection*{Spheroids}
\begin{figure*}
\centering
\includegraphics[width=.95\textwidth]{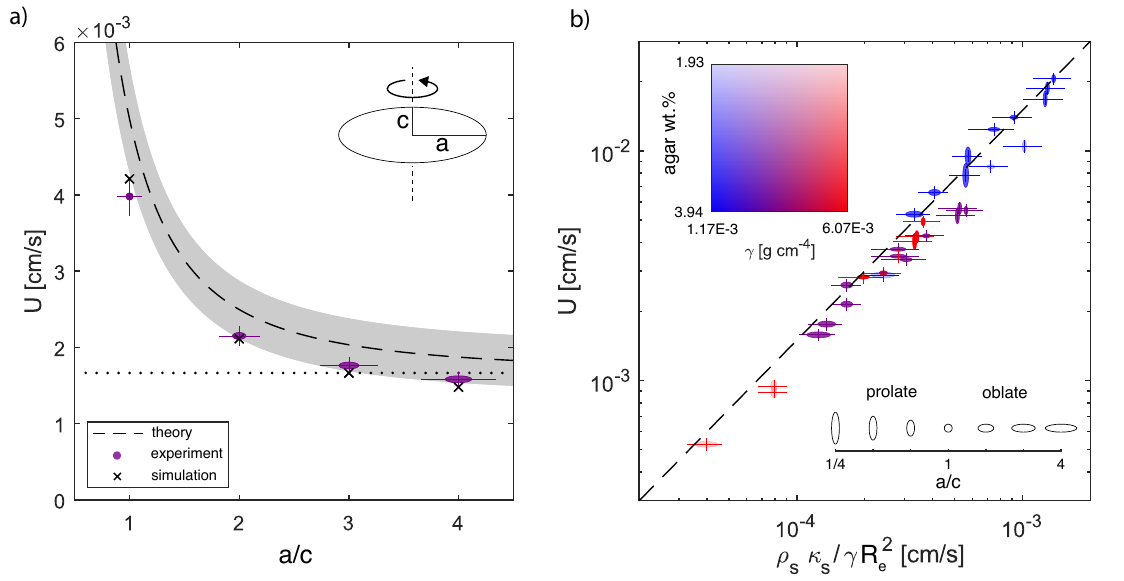}
\caption{a) Dimensional plot of settling velocity $U$ versus aspect ratio $a/c$ for oblate agar spheroids, where $c$ represents length of the semi-axis aligned with the axis of rotational symmetry and $a$ represents the length of the other two semi-axes. Here, $c = .3$ cm, $a = .3 - 1.2$ cm, agar wt.\% = 3.4, $\rho_s = 1.0\times10^{-2}$ g cm$^{-3}$, $\kappa_p = 1.0\times10^{5} $ cm$^2$ s$^{-1}$, and $\gamma = 3.5\times10^{-3}$ g cm$^{-4}$. The theoretical prediction (Equations \ref{eq:exactsol} and \ref{eq:ellipsoidshapefactor}) is represented by a black dashed line with propagated uncertainty associated with uncertainty in measured parameters shown in grey. The predicted asymptotic value in the limit of large $a/c$ is shown as a dotted line. Results from numerical simulations are depicted by a black $\times$. Experimental results are purple, corresponding to their percentage of agar by weight and density gradient as in panel b), with error bars representing uncertainty in $a/c$ and $U$. b) Non-dimensional collapse of experimentally measured settling velocity $U$ of non-spherical spheroids versus a combination of measured parameters that scale with the theoretical prediction for settling velocity (Equations \ref{eq:exactsol} and \ref{eq:ellipsoidshapefactor}, represented by the black dashed line). Marker size indicates relative spheroid size, marker shape indicates the aspect ratio, color indicates strength of gradient, and opacity represents the agar wt.\% as shown in the legends. Raw data included in SI dataset S1 \& S2.} 
\label{fig:ShapeDependenceDimensional}
\end{figure*}

\begin{figure}
\centering
\includegraphics[width=.5\linewidth]{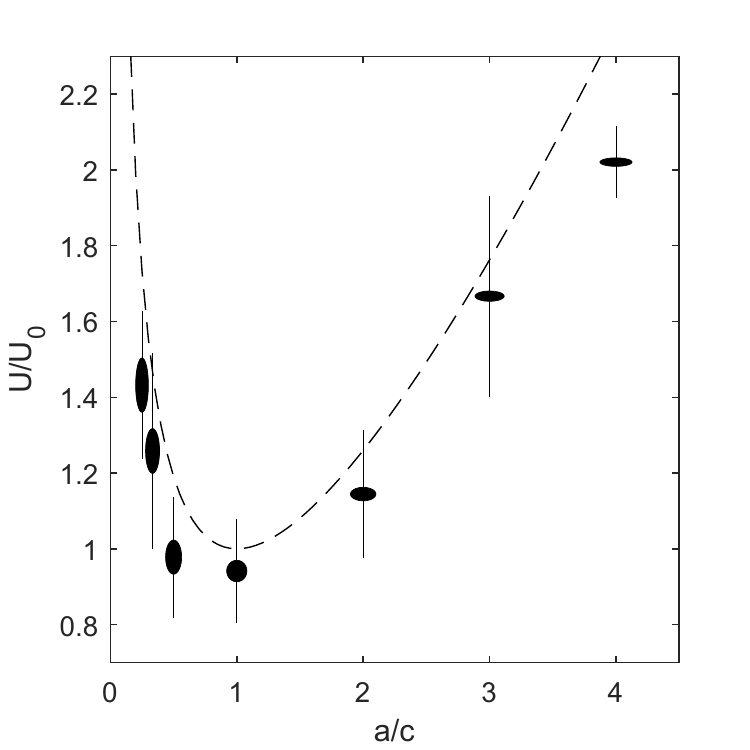}
\caption{Mean settling velocity $U$ for spheroids (data from Figures \ref{fig:SpheresDimensional}(b) and \ref{fig:ShapeDependenceDimensional}(b)) normalized by velocity $U_0$ predicted for a sphere of the same volume, as a function of aspect ratio $a/c$. Error bars represent the standard deviation over all trials for a given aspect ratio. Theoretical prediction is depicted by a black dashed line. All particles of the same aspect ratio $a/c$ are summarized by the mean and standard deviation of this normalized velocity measure over all parameters. Raw data included in SI dataset S1.}
\label{fig:ShapeFactor}
\end{figure}
Although spheres are a natural starting point and thus represent the overwhelming focus of the literature thus far, there are many cases where the shape is highly non-spherical, for example in thin aggregate discs as often observed in marine snow. As discussed prior, one advantage of the diffusion-limited limit is the ease of extending the prediction to bodies of arbitrary shape.

In Figure \ref{fig:ShapeDependenceDimensional}(a), we consider the settling of an oblate spheroid whose vertical semiaxis $c$, corresponding to its axis of rotational symmetry, is fixed at 0.3 cm, while the horizontal semiaxis $a=b$ is varied from 0.3 to 1.2 cm, spanning a range of aspect ratios $a/c$ from 1 to 4. The measured settling velocities are compared to the prediction of Equation \ref{eq:exactsol} using the effective radius defined in Equation \ref{eq:ellipsoidshapefactor}. As in Figure \ref{fig:SpheresDimensional}(a), we observe a slight but consistent overprediction of the settling velocity that may be due to the presence of the convective boundary layer just outside of the solid particle boundary. The predictions from the full numerical simulations again show excellent agreement with the experimentally measured velocities. For large aspect ratio, the oblate spheroids asymptotically approach a constant settling velocity, with the predicted asymptotic value pictured as a dotted line (see Supplementary Information for details). In Figure \ref{fig:ShapeDependenceDimensional}(b), aggregate settling data for non-spherical spheroids is presented with aspect ratios $a/c$ ranging from 1/4 to 4 and over a range of agar percentages and linear stratifications. Good overall collapse to the theoretical prediction is found, with a maximum error of 35\% and a mean error of 15\%, moving 13\% slower than predicted on average.

As diffusive transport is inherently a surface-dominated phenomenon, it is anticipated that for a fixed particle volume, a sphere will settle at the slowest rate as it represents the shape of minimal surface area. In Figure \ref{fig:ShapeFactor}, we replot and summarize the data from Figures \ref{fig:SpheresDimensional}(b) and \ref{fig:ShapeDependenceDimensional}(b) as a function of particle aspect ratio $a/c$. The settling speeds are normalized by the theoretical settling speed $U_{0}= \frac{15 \kappa_p \rho_s}{\gamma R_{0}^2}$ of an otherwise equivalent sphere of the same volume (corresponding to $R_{0} = (a^2c)^{1/3}$). All particles of the same aspect ratio $a/c$ are summarized by the mean and standard deviation of this normalized velocity measure over all parameters. Consistent with the intuitive argument, a minimum normalized velocity is predicted when the object is spherical, and the overall trend is well supported by experiments.

\begin{table}

\centering
\begin{tabular}{ c  c c  }

\begin{minipage}{.08\textwidth}
      \includegraphics[width=\textwidth]{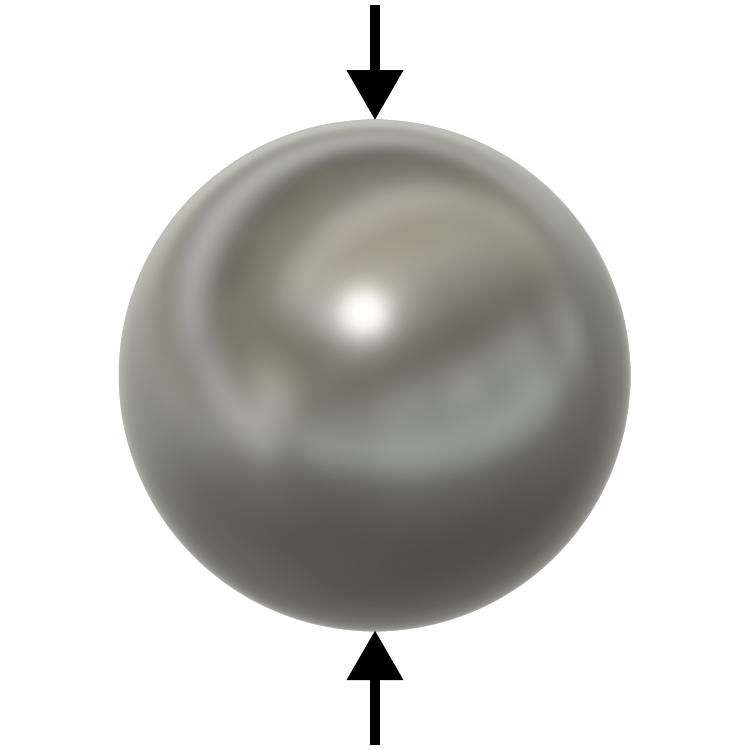}
    \end{minipage}
 & sphere & $1$  \\

 \begin{minipage}{.08\textwidth}
      \includegraphics[width=\textwidth]{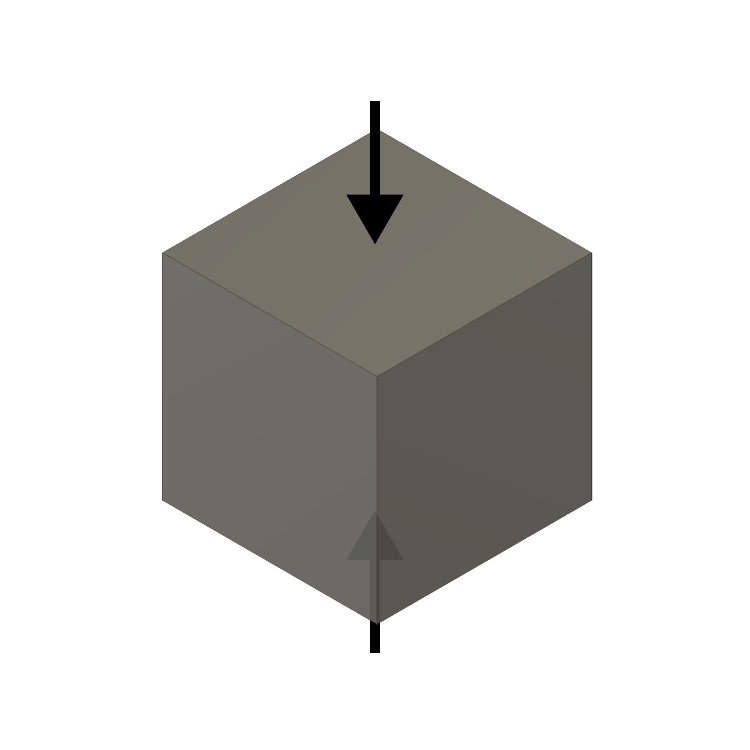}
    \end{minipage}
 & cube &$ \approx0.826 $\\

  \begin{minipage}{.08\textwidth}
      \includegraphics[width=\textwidth]{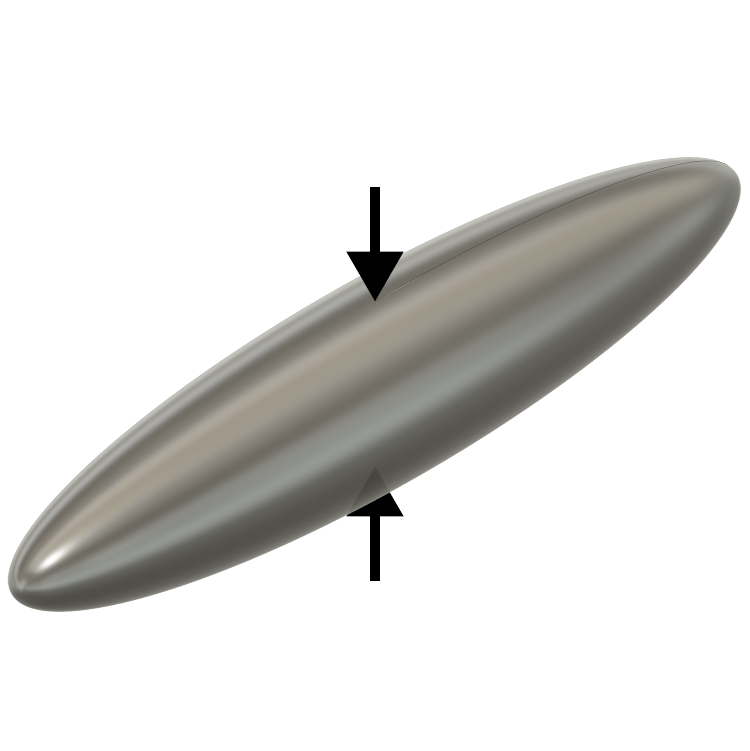}
    \end{minipage}
 &   long prolate ellipsoid   & $2/3$ \\

     \begin{minipage}{.08\textwidth}
      \includegraphics[width=\textwidth]{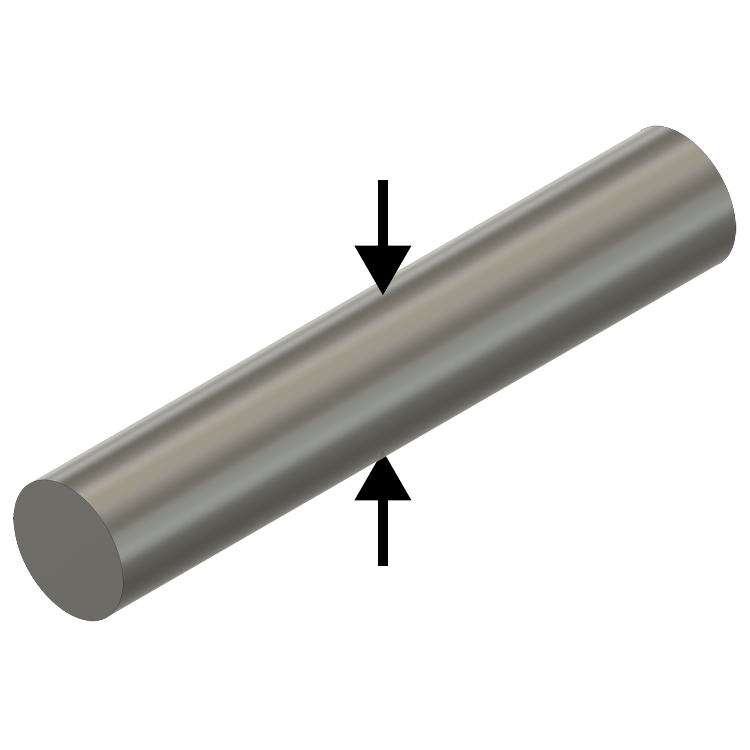}
    \end{minipage}
 &   long cylindrical rod   & $8/15$ \\

       \begin{minipage}{.08\textwidth}
      \includegraphics[width=\textwidth]{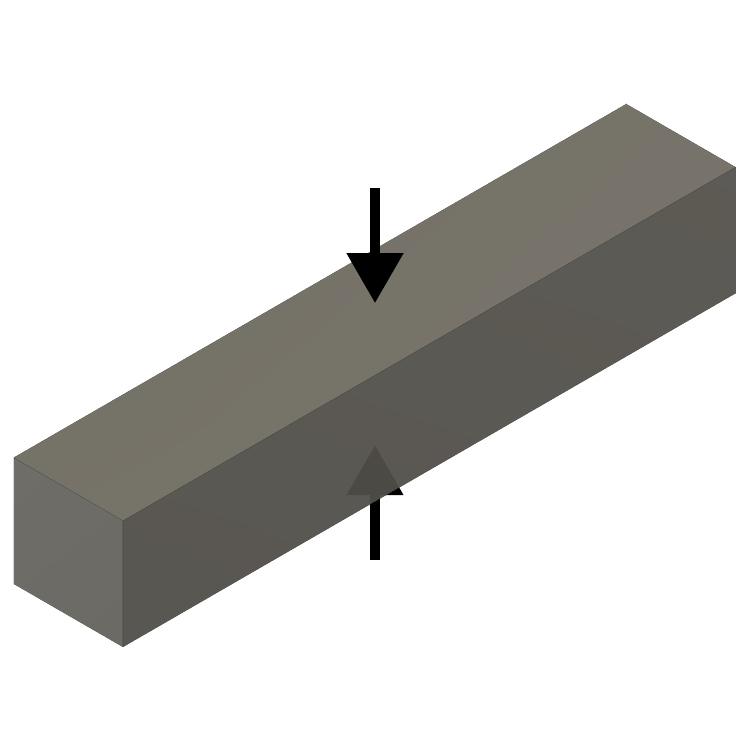}
    \end{minipage}
 & long square rod & $\approx0.474$ \\

   \begin{minipage}{.08\textwidth}
      \includegraphics[width=\textwidth]{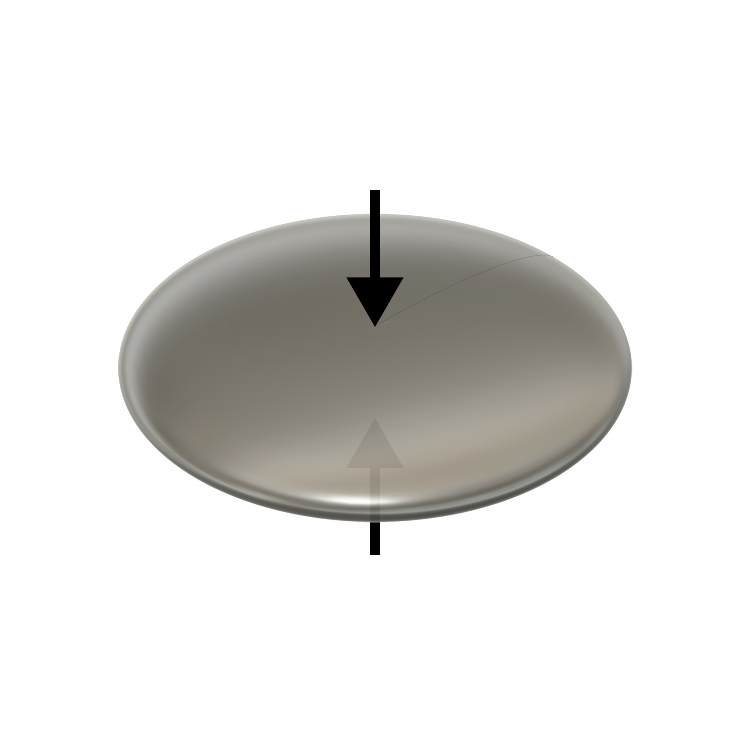}
    \end{minipage}
 &   \begin{minipage}{.3\textwidth}thin oblate ellipsoid\end{minipage}   & $1/3$ \\

      \begin{minipage}{.08\textwidth}
      \includegraphics[width=\textwidth]{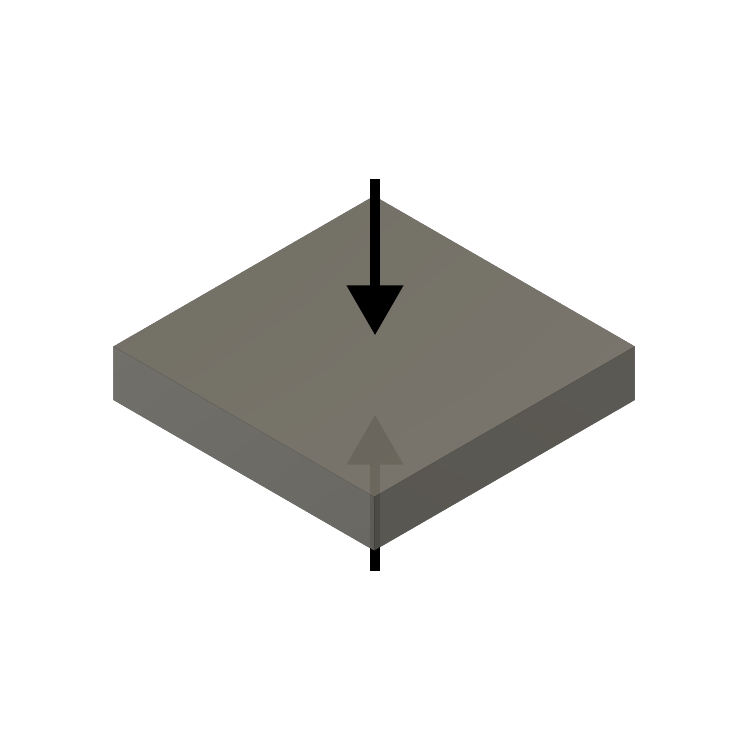}
    \end{minipage}
 & thin uniform sheet  & $1/5$ \\
\hline
\end{tabular}
\caption{Diffusion-limited settling velocity ratio $U_{\gamma}/U_{d}$ for shapes with thin characteristic dimension $d$ (indicated by arrows) relative to a sphere of the same diameter ($U_d = \frac{60 \kappa_p \rho_{s}}{\gamma d^{2}}$). Derivations and exact results for more geometries are included in the Supplementary Information.}

\label{table:shapes}
\end{table}

\subsection*{Slender bodies}
Our formula also allows for predictions of the settling of asymptotically slender bodies in the diffusion-limited regime (although only tested to aspect ratio 4). The settling speed for objects characterized by a thin dimension $d$ relative to a sphere of the same diameter is summarized in Table \ref{table:shapes}. For a pancake-like oblate spheroid with long horizontal semiaxes of length $a=b=l$ and short vertical semiaxis $c=d/2 \ll l$, the settling speed asymptotically approaches 1/3 that of a sphere of diameter $d$. For a fiber-like prolate spheroid with long vertical semiaxis of length $c=l$ and short horizontal semiaxes of length $a=b=d/2 \ll l$, the settling speed approaches 2/3 that of a sphere of diameter $d$.  Furthermore, for the case of a long cylindrical fiber of diameter $d$ or a thin uniform sheet of thickness of $d$, it can be shown that the settling speed approaches 8/15 and 1/5 that of a sphere of diameter $d$, respectively. In all of these extreme cases, the settling velocity simply scales as $U \sim \frac{\rho_s \kappa_p}{\gamma d^2}$, where the dependence on the long dimension is lost, as the relevant diffusion timescale is driven by the small dimension. It is anticipated that any high-aspect ratio body will follow a similar scaling, with the smallest dimension dictating the settling speed. A discussion regarding these limits and criteria for validity of the diffusion-limited regime can be found in the Supplementary Information.

\subsection*{Non-linear stratifications}
\label{section:densityFromTrajectory}

Although our primary results focus on steady settling in linear stratifications, for systems where the local background density gradient changes slowly relative to the particle diffusion timescale, these results are also applicable to non-linear stratifications. A particle settling at the diffusion-limited velocity $U_{\gamma}$ travels a distance $\rho_{s}/\gamma$ over one particle diffusion time ($t_{\gamma} = R^2/\kappa$). Thus for the prior analysis to remain valid, the local density gradient should be approximately constant over such a length scale. From this argument, one can arrive at a condition for the second derivative of the density profile:
\begin{equation}
\label{eq:quasisteadycondition}
    \left|\frac{d^2}{d z^2}\rho_{b}(z)\right| \ll \frac{\gamma^2}{\rho_{s}}.
\end{equation}
More details regarding this derivation can be found in the Supplementary Information. Thus, in the regime where fluid drag is negligible (large $Ra$) and in regions where the above criterion is satisfied, we anticipate the local depth-dependent settling velocity to be described by the natural extension of Equation \ref{eq:ugamma}:
\begin{equation}
\label{eq:ulocal}
U(Z) = \frac{15 \kappa_p \rho_{s}}{\gamma(Z) R^{2}}. 
\end{equation}

\subsection*{Reconstruction of density profile from particle trajectory}
\label{sec:reconstruction}

\begin{figure*}
\centering
\includegraphics[width=\textwidth]{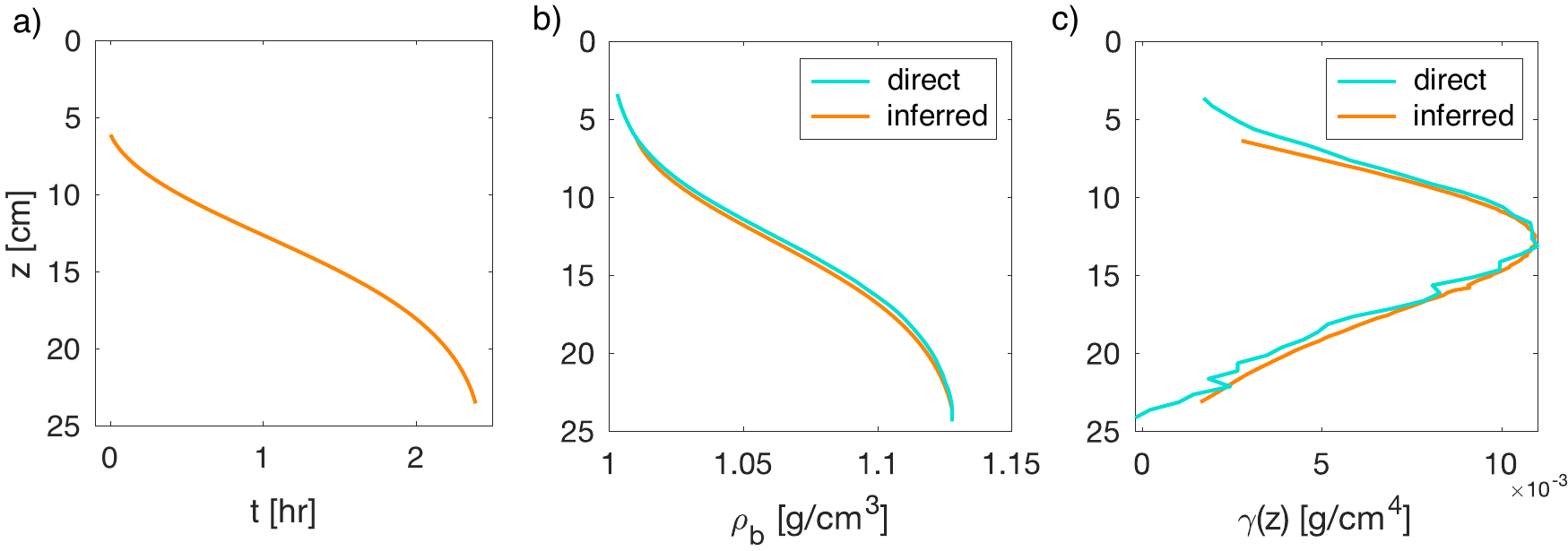}
\caption{a) Particle position versus time from experiments for a sphere ($R = 3$ mm, $3.4$ wt.\% agar). b) Comparison of direct measurement of density profile (orange) and measurement inferred from particle trajectory (cyan). c) Comparison of direct measurement of density gradient (orange) and measurement inferred from particle trajectory (cyan). Raw data included in SI dataset S3.}
\label{fig:DensityMeasurement}
\end{figure*}

The condition that the diffusion-induced settling velocity is satisfied locally for an arbitrary density profile (Equation \ref{eq:ulocal}) further implies that the local background density rate of change ($\frac{d\rho_b}{dt}=U\gamma$) is constant in time. This consequence can be seen readily from the relation
\begin{equation}
\label{eq:balance1}
\frac{d\rho_b}{dt} = U(Z(t))\gamma(Z(t)) = \frac{15 \rho_s \kappa_p}{R_e^2},
\end{equation}
 which depends only on the particle's extra solid density, diffusivity, and radius. For a given particle, the value of this rate of change ($\psi=\frac{15 \rho_s \kappa_p}{R_e^2}$) can be estimated directly using the particle size and material parameters. Alternatively, if one knows the fluid density at the initial ($\rho_i$) and final ($\rho_f$) particle positions, along with the total transit time ($t_r$), $\psi$ can be computed directly as $\psi=\frac{\rho_f - \rho_i}{t_r}$. Given knowledge of this density rate of change and the initial density, one can then reconstruct the density profile directly from the particle trajectory, assuming the background stratification itself is evolving slowly relative to the measurement timescale. In practice, this reconstruction is done by exploiting the linear relationship between time and density ($\rho_b(t) = \rho_b(0) + \psi t$) to reparameterize the position as a function of density. This framework allows for a simple, cheap, highly resolved, and minimally invasive method for density profile reconstruction using only two density measurements and a camera. 
 
 An example experiment, as depicted in Figure \ref{fig:DensityMeasurement},
involves dropping a small agar particle in a tank that was stratified with a hyperbolic tangent density profile using the method described in the Methods section. As can be seen in Figure \ref{fig:DensityMeasurement}(a), the particle does not settle at a constant speed in this case, due to the fact that the background gradient is no longer constant. Using the initial density, final density, and measured transit time, one can reconstruct the density profile as shown in Figure \ref{fig:DensityMeasurement}(b). This result shows excellent agreement with the density profile measured directly using a conductivity probe mounted to a motorized linear positioning stage. One can also faithfully estimate the local density gradient using such data, as shown in Figure \ref{fig:DensityMeasurement}(c).

As a related aside, should the sphere size and properties be known independently, one can calculate the transit time $t_{r}$ across a pycnocline in the diffusion-limited regime as
\begin{equation}
\label{eq:transit}
    t_{r} = \frac{R_e^2}{15 \kappa_p \rho_s}(\rho_{f} - \rho_{i}).
\end{equation}
Notably this time is predicted to depend only on the particle properties (embodied in $\psi$) and the traversed density difference ($\rho_{f} - \rho_{i}$) but is independent of the overall thickness of the pycnocline and the details of its functional form. As discussed prior, this result should hold provided the particle is in the diffusion-limited regime and when Equation \ref{eq:quasisteadycondition} is satisfied. The actual transit time is anticipated to deviate from this prediction when the stratification is sharp relative to the particle size, for instance. Although not explicitly tested in this work, the square dependence of particle radius on the transit time is consistent with prior laboratory measurements and analyses for the case of spheres \cite{li2003,kindler2010, camassa2013}.  Our result goes beyond the scaling and also provides a prediction for bodies of of arbitrary shape via the suitably defined effective radius $R_e$.
 
\section*{Discussion}
In this work, we have derived a simple analytical formula for the diffusion-limited settling speed of highly porous particles in a linear background stratification. The assumption of a linear density gradient is in contrast with most previous laboratory studies and simulations that consider variations in the density gradient on the order of the particle size. However, in many systems that motivate the current investigation, there is a vast separation of scales between the particle size and background density variations, motivating the study of particles in a locally linear gradient. Diffusion-limited settling corresponds to scenarios where mass exchange between the particle and background fluid is the dominant mechanism determining the particle position, with hydrodynamic drag being negligible. For the case of simple Stokes drag, this regime can be delineated by a suitably defined Rayleigh number. Despite the assumptions and simplicity of the final result, the prediction shows good agreement with new experiments on spherical hydrogel particles for a range of particle radii, compositions, and background stratifications. On average, the results slightly overpredict the measured settling speeds, which we expect is predominantly due to the neglect of the small density boundary layer surrounding the particle that additionally buffers mass transport into the sphere.

In oceans, salt lakes, and estuaries, vertical salinity stratifications are commonly on the order of $\gamma\approx 10^{-5} - 10^{-9}$ g/cm$^4$ \cite{ahmerkamp2022,IAPdata}. Thus assuming a settling law of the form of equation \ref{eq:Ufull}, diffusion-limited settling represents a dominant retention mechanism for particles with $R \gtrsim 0.16 - 1.6$ cm (given $\mu = 0.01$ g cm$^{-1}$, $g = 981$ cm s${^{-2}}$, $\kappa_p = 10^{-5}$ cm$^2$ s${^{-1}}$). Despite the idealized scenario considered herein, it appears plausible that such a mechanism may play an important role in environmental systems, with the effect being most relevant for larger aggregates. As particle size has important implications for bioavailability and microbial respiration \cite{JOHNSON1992}, the differential settling due to size imposed by this mechanism may be relevant to these chemical and ecological processes.

A distinguishing feature of the diffusion-limited regime involves the dependence on size and shape. For a given solid density, the settling velocity of a sphere in a uniform viscous fluid scales with the square of the radius, as described by Stokes settling law. In the diffusion-limited regime, the settling velocity for a sphere scales \textit{inversely} with the square of the radius, as previously proposed and documented for diffusion-limited retention \cite{li2003,kindler2010,camassa2013}. Although more complex settling behaviors for non-porous particles have been investigated to include effects of density stratification \cite{zvirin1975, yick2009, more2023}, these results retain the monotonic increase in particle velocity dependence on size, in contrast with the current work.

Although a sphere gives the simplest geometrical representation of a particle, in many motivating systems of interest particles are highly non-spherical. Essentially all prior related laboratory and modeling studies have restricted attention to spherical particles. The role of shape has been identified as essential for understanding particle transport in the atmosphere and oceans \cite{tatsii2023, BYRON2019, sutherland2023}. In this work, we have also experimentally validated an analogous prediction for spheroids and outlined a mathematical framework for computing the diffusion-limited settling velocity of an arbitrary geometry. As a boundary-flux driven phenomenon, for a given particle volume, a sphere settles with the slowest speed, as it represents the shape with minimal surface area for a given volume. Our model also provides insight into the settling dynamics of particles with very large aspect ratios that may be harder to access experimentally.

We have also derived a criterion under which our main results can be readily applied to non-linear stratifications. Through this analysis, we developed a simple and accessible method for using particle trajectories settling in the diffusion-limited regime to reconstruct highly resolved background density profiles.

In our model, particles are considered as homogeneous diffusive materials which admit an effective diffusivity $\kappa_p$ and carry an additional component of density $\rho_s$ due to the presence of solid material. In generality, porous materials may contain regions which are inaccessible to the solute, modeled as a solid volume fraction. The consideration of this effect may be added through a partition function \cite{PHILLIPS1990363}, causing a discontinuity in the averaged solute concentration. This can be accounted for in the framework of our study by a prefactor in the settling velocity expression which is captured when measuring $\Psi$ directly, as in the previous section (for details see Supplementary Information). Other considerations, including tortuosity of the medium, hydrodynamic effects, fluid flow through the porous medium, and other complex phenomena relevant to diffusion in porous materials may invalidate the assumption of homogeneous and isotropic diffusivity considered here. Nevertheless, we generally expect our model's assumptions to remain valid in the limit of low solid volume fraction and small pore size.

\section*{Methods}
\label{section:methods}
\subsection*{Fluid preparation}
To create the background density stratification, we use two programmable pulsatile pumps driven by stepper motors (Kamoer PHM400-ST3B25) that are controlled by an Arduino Uno running the AccelStepper library which sends step and direction information to two TMC2209 drivers. One pump supplies dense saline solution, and the other pump supplies less dense saline solution or fresh deionized water. These two input streams are combined through a 3D-printed Y junction and fed into the experiment tank through a single tube. The experiment tank (REPTIZOO B07CV8L7BK) is made of glass with interior dimensions of 19 cm x 19 cm x 28 cm width, depth, and height, respectively. The tank is rinsed with deionized water and dried before pouring the stratification. A diffuser, consisting of a sponge surrounded by a foam border, floats at the water surface and allows for incident saline solution to not disturb the density-stratified layer structure below.

To prepare the saline solution, we first rinse sodium chloride (Morton Pure and Natural Water Softener Crystals) with deionized, reverse-osmosis filtered water. To avoid introducing dissolved gases into the saline solution during mixing, the deionized water is boiled before pouring the stratification. The sodium chloride is dissolved and the stratification is poured typically within one hour of boiling. A small sample of both the dense and fresh reservoirs is set aside and allowed to cool before measuring the density using an Anton Parr DMA35 densitometer.

To confirm the accuracy of our stratification pouring method, we use a Mettler Toledo SevenCompact S230 conductivity meter and InLab 731-ISM probe, which collects conductivity measurements as a function of depth. The depth is controlled automatically through a stepper driver and motorized stage (HoCenWay DM556, Befenbay BE069-4), triggering a camera to record the measured conductivity. The relation between conductivity and density was calibrated by preparing 22 saline solutions whose density and conductivity was measured. Direct measurements of density gradients for prepared linear stratifications were within the reported uncertainty of 10\% from the predicted value.

\subsection*{Particle preparation}
To prepare the agar particles, agar powder (Acros Organics 400405000) was combined with boiling, deionized water at a known weight ratio (typically 2-4 wt.\%) and mixed using an immersion blender until well-mixed, typically around 30 seconds. The particles are cast using a 2-part 3D-printed mold. The mold is cleaned and dried, then prepared by clamping the upper and lower section together using spring clamps. A 5 ml syringe is filled with hot agar solution before a 20G needle is attached. The hot agar solution is injected into the mold through a small hole at the top. The solution is inspected to confirm the absence of large bubbles. After allowing for the cast agar to solidify, typically for 15-30 minutes, the clamps are removed and the mold is carefully separated. Particles are placed in deionized water and allowed to rest, typically overnight, before being used in experiments. The particle size uncertainty was estimated from images as $\pm .025$ cm.

\subsection*{Particle characterization}
To measure the particle density, we use a force balance method. A small measuring stage is held by a thin wire and a long horizontal support arm whose base is rested on a scale (U.S. Solid USS-DBS5). The measurement stage is immersed in a small container of deionized water which is placed on a second scale. The scales are zeroed, then a small sample of particles is placed on the measuring stage. Due to this configuration, the scale on which the water cup is resting measures the weight corresponding to the volume of water displaced by the particle sample. The scale which supports the horizontal support arm that the measuring stage is attached to measures only the extra weight that is due to the excess density of the particle sample (the component of the density that exceeds the deionized water in which they are immersed).

To measure the diffusivity of the agar particles, we similarly used particles immersed in a solution of a known density. Typically, a selection of five spheres of different diameter were submerged in a uniform density saline solution and, although they are initially positively buoyant, held underwater using a laser cut acrylic frame and 200 $\mu$m diameter nylon monofilament structure. The time at which the particle begins to settle is recorded, and it is assumed at this time that the mean density of the particle is equal to that of the background saline solution. To estimate the effective diffusivity, we then employ a mathematical model, assuming that the sphere can be approximated as a material with a uniform diffusivity whose external boundary condition for density is set by that of the uniform background saline solution density, neglecting potential variations in the external density field. The transient density of the sphere is modeled using an analytic solution for the salt concentration distribution $\sum _{n=1}^{\infty } \frac{6 \Delta \rho}{ \pi ^2 n^2}  e^{-\pi ^2 n^2 \kappa_{p} t / R^2}$ which is truncated at 1000 terms. We use a bisection search method to find the effective diffusivity of salt in the particle which corresponds to the observed settling time given the above assumptions.

Due to this measurement process’ dependence on the extra solid density measurement, the uncertainty in the diffusivity value is highly correlated with the measured extra density. For this reason, we employ a Monte Carlo method to estimate the variance of the combined product of extra density and diffusivity as used in the empirical settling velocity estimate. First, we collect a set of excess density measurements and find the mean and variance of this distribution. Then, we collect a set of settling times from the diffusivity measurement procedure described above and record the normalized settling time, which is given by the settling time divided by the sphere radius squared. From this, we can deduce the mean and variance of the normalized settling time distribution. At this step, we create synthetic data by modeling the excess density and normalized settling time as normally distributed random variables with mean and variance as measured from experiment. We randomly sample an extra density and a normalized settling time, then calculate the effective diffusivity and the product of the extra density and effective diffusivity. We repeat this procedure for 10,000 sample pairs and then calculate the mean and variance of the generated population of products of extra density and effective diffusivity. The standard deviation due to the extra density and diffusivity was found to be 1.8\% for an agar weight ratio of 3.44\% and was taken as typical for all agar weight ratios.

\subsection*{Particle kinematics}
To measure the particle position in experiments, a camera (Nikon D850 equipped with a Nikon Nikkor 105mm lens) views the experiment tank from the side and records images typically every 20 seconds, resolved at 53 $\mu$m per pixel. The tank is fitted with a dark colored background and illuminated from the side, causing the particles to appear bright relative to the image background. The image is binarized, and the center of the largest bright region is taken as the particle center. To deduce the settling velocity, the center position as a function of time is linearly approximated by least squares over a 5 cm region near the center of the tank. The conversion from image to real space coordinates involves measuring a length scale directly from the image of the tank. Due to parallax, this scale varies for a particle at the front or rear of the tank. The nominal scaling factor is taken as the mean of these two limiting scaling factors, and the uncertainty bounds are taken as the minimum and maximum values ($\pm 6.3\%$).

\subsection*{Simulations}
The fluid-solute system is modeled in COMSOL 5.6 using the $\textit{Laminar Flow}$ and $\textit{Transport of Diluted Species}$ packages as a 2D axisymmetric geometry. The simulation is performed in the frame of the sphere or spheroid, where the mesh is fixed in time. The linear solute concentration is prescribed at the inlet, lateral boundary, and outlet. To simulate the process of settling in the frame of the sphere, the concentration boundary conditions are advected in time in conjunction with the imposed velocity. A uniform velocity is prescribed at the inlet, and the static pressure is prescribed at the outlet. The velocity at the inlet is linearly ramped to a constant velocity $U$ over 1/5 of the diffusion time $t_{\gamma}$. Simulations are run for $2 t_{\gamma}$, until the density perturbation relative to the background is constant, utilizing a BDF time-stepping scheme with maximum stepsize $t_{\gamma}/100$. The fluid parameters are matched with saline water, with dynamic viscosity $\mu = .011219$ g cm$^{-1}$ s$^{-1}$, background density $.997$ g cm$^{-3}$, and solute diffusivity in the fluid $1.5 \times 10^{-5}$ cm$^2$ s${^{-1}}$. This model is coupled to a solid particle region with a no-slip boundary condition for the fluid velocity, continuity of solute concentration, and continuity of solute flux. Inside the solid, the solute diffusivity $\kappa_p$ is determined from experiment as described in the Particle Characterization section.

The cylindrical domain radius is chosen to be 10 times the horizontal semiaxis, and the height is 20 times the vertical semiaxis. The mesh is generated with a minimum element size of 0.0001 times the horizontal semiaxis, with a maximum element growth rate of 1.0005 and a curvature factor of 0.08. To resolve the solute and velocity boundary layer, a boundary layer mesh is prescribed along the spheroid surface and upper vertical axis with $\textit{Number of boundary layers}$ equal to 10 and $\textit{Boundary layer stretching factor}$ equal to 0.8.

At steady-state, the stress on the solid boundary is integrated to calculate the total force acting on the particle. For the particle to have reached a force equilibrium in this configuration, the extra body force due to the solid density $\rho_s$ must offset this value. Thus in essence we impose a settling velocity, which then defines the solid density assuming equilibrium (opposite of the experiments). To compare with specific experimental trials, we iterate this procedure by updating the simulated settling velocity until the force on the particle is equal to the experimentally measured particle solid density $\rho_{s}$. This iteration procedure is accelerated by assuming the diffusion-limited velocity relationship between $\rho_s$ and $U_{\gamma}$ as described in Results to update the velocity used in the simulation. This iteration scheme is repeated until the calculated solid density from simulations matches the experimentally measured $\rho_s$ within 1.2\% tolerance.

\section*{Acknowledgements}
This work was partially funded by NSF DMS-1909521, NSF DMS-1910824, NSF DMS-2308063, ONR N00014-18-1-2490, and ONR N00014-23-1-2478. We would like to thank Professors Monica Martinez-Wilhelmus and Roberto Zenit for loaned equipment, Rebecca Rosen for support with preliminary experiments, and John Antolik for discussions.

\section*{Author Contributions}
R.H. proposed research. R.H. and D.M.H. designed research. R.H. performed research. R.H., R.C., R.M.M., and D.M.H. discussed and interpreted results. R.H. and D.M.H. wrote the paper. R.H., R.C., R.M.M., and D.M.H. edited the paper.

\section*{Competing Interests}
The authors declare no competing interests.
\bibliographystyle{abbrv}
\bibliography{main}

\begin{thebibliography}{10}

\bibitem{abaid2004}
N.~Abaid, D.~Adalsteinsson, A.~Agyapong, and R.~M. McLaughlin.
\newblock {An internal splash: Levitation of falling spheres in stratified fluids}.
\newblock {\em Physics of Fluids}, 16(5):1567--1580, 05 2004.

\bibitem{ahmerkamp2022}
S.~Ahmerkamp, B.~Liu, K.~Kindler, J.~Maerz, R.~Stocker, M.~Kuypers, and A.~Khalili.
\newblock Settling of highly porous and impermeable particles in linear stratification: implications for marine aggregates.
\newblock {\em Journal of Fluid Mechanics}, 931:A9, 2022.

\bibitem{alldredge1988}
A.~L. Alldredge and C.~Gotschalk.
\newblock In situ settling behavior of marine snow.
\newblock {\em Limnology and Oceanography}, 33(3):339--351, 1988.

\bibitem{alldredge1987}
A.~L. Alldredge, C.~C. Gotschalk, and S.~MacIntyre.
\newblock Evidence for sustained residence of macrocrustacean fecal pellets in surface waters off southern california.
\newblock {\em Deep Sea Research Part A. Oceanographic Research Papers}, 34(9):1641--1652, 1987.

\bibitem{BYRON2019}
M.~L. Byron, Y.~Tao, I.~A. Houghton, and E.~A. Variano.
\newblock Slip velocity of large low-aspect-ratio cylinders in homogeneous isotropic turbulence.
\newblock {\em International Journal of Multiphase Flow}, 121:103120, 2019.

\bibitem{Camassa2022}
R.~Camassa, L.~Ding, R.~M. McLaughlin, R.~Overman, R.~Parker, and A.~Vaidya.
\newblock {\em Critical Density Triplets for the Arrestment of a Sphere Falling in a Sharply Stratified Fluid}, pages 69--91.
\newblock Springer International Publishing, Cham, 2022.

\bibitem{camassa2010}
R.~Camassa, C.~Falcon, J.~Lin, R.~M. McLaughlin, and N.~Mykins.
\newblock A first-principle predictive theory for a sphere falling through sharply stratified fluid at low reynolds number.
\newblock {\em Journal of Fluid Mechanics}, 664:436–465, 2010.

\bibitem{camassa2009}
R.~Camassa, C.~Falcon, J.~Lin, R.~M. McLaughlin, and R.~Parker.
\newblock {Prolonged residence times for particles settling through stratified miscible fluids in the Stokes regime}.
\newblock {\em Physics of Fluids}, 21(3):031702, 03 2009.

\bibitem{camassa2019}
R.~Camassa, D.~M. Harris, R.~Hunt, Z.~Kilic, and R.~M. McLaughlin.
\newblock A first-principle mechanism for particulate aggregation and self-assembly in stratified fluids.
\newblock {\em Nature Communications}, 10(1):5804, 2019.

\bibitem{camassa2013}
R.~Camassa, S.~Khatri, R.~M. McLaughlin, J.~C. Prairie, B.~L. White, and S.~Yu.
\newblock {Retention and entrainment effects: Experiments and theory for porous spheres settling in sharply stratified fluids}.
\newblock {\em Physics of Fluids}, 25(8):081701, 08 2013.

\bibitem{fazey2016}
F.~M. Fazey and P.~G. Ryan.
\newblock Biofouling on buoyant marine plastics: An experimental study into the effect of size on surface longevity.
\newblock {\em Environmental Pollution}, 210:354--360, 2016.

\bibitem{JOHNSON1992}
B.~D. Johnson and P.~E. Kepkay.
\newblock Colloid transport and bacterial utilization of oceanic doc.
\newblock {\em Deep Sea Research Part A. Oceanographic Research Papers}, 39(5):855--869, 1992.

\bibitem{kindler2010}
K.~Kindler, A.~Khalili, and R.~Stocker.
\newblock Diffusion-limited retention of porous particles at density interfaces.
\newblock {\em Proceedings of the National Academy of Sciences}, 107(51):22163--22168, 2010.

\bibitem{IAPdata}
G.~Li, L.~Cheng, Y.~Pan, G.~Wang, H.~Liu, J.~Zhu, B.~Zhang, H.~Ren, and X.~Wang.
\newblock A global gridded ocean salinity dataset with 0.5° horizontal resolution since 1960 for the upper 2000 m.
\newblock {\em Frontiers in Marine Science}, 10, 2023.

\bibitem{li2003}
X.-Y. Li, Y.~Yuan, and H.-W. Wang.
\newblock Hydrodynamics of biological aggregates of different sludge ages: An insight into the mass transport mechanisms of bioaggregates.
\newblock {\em Environmental Science \& Technology}, 37(2):292--299, 2003.

\bibitem{liu2022}
S.~Liu, Y.~Huang, D.~Luo, X.~Wang, Z.~Wang, X.~Ji, Z.~Chen, R.~A. Dahlgren, M.~Zhang, and X.~Shang.
\newblock Integrated effects of polymer type, size and shape on the sinking dynamics of biofouled microplastics.
\newblock {\em Water Research}, 220:118656, 2022.

\bibitem{macintyre1995}
S.~MacIntyre, A.~L. Alldredge, and C.~C. Gotschalk.
\newblock Accumulation of marines now at density discontinuities in the water column.
\newblock {\em Limnology and Oceanography}, 40(3):449--468, 1995.

\bibitem{mehaddi2018}
R.~Mehaddi, F.~Candelier, and B.~Mehlig.
\newblock Inertial drag on a sphere settling in a stratified fluid.
\newblock {\em Journal of Fluid Mechanics}, 855:1074–1087, 2018.

\bibitem{more2023}
R.~V. More and A.~M. Ardekani.
\newblock Motion in stratified fluids.
\newblock {\em Annual Review of Fluid Mechanics}, 55(Volume 55, 2023):157--192, 2023.

\bibitem{ostrach1953}
S.~Ostrach.
\newblock An analysis of laminar free-convection flow and heat transfer about a flat plate paralled to the direction of the generating body force.
\newblock In {\em No. NACA-TR-1111}, 1953.

\bibitem{panah2017}
M.~Panah, F.~Blanchette, and S.~Khatri.
\newblock Simulations of a porous particle settling in a density-stratified ambient fluid.
\newblock {\em Phys. Rev. Fluids}, 2:114303, Nov 2017.

\bibitem{PHILLIPS1990363}
R.~J. Phillips, W.~M. Deen, and J.~F. Brady.
\newblock Hindered transport in fibrous membranes and gels: Effect of solute size and fiber configuration.
\newblock {\em Journal of Colloid and Interface Science}, 139(2):363--373, 1990.

\bibitem{prarie2013}
J.~Prairie, K.~Ziervogel, C.~Arnosti, R.~Camassa, C.~Falcon, S.~Khatri, R.~McLaughlin, B.~White, and S.~Yu.
\newblock Delayed settling of marine snow at sharp density transitions driven by fluid entrainment and diffusion-limited retention.
\newblock {\em Marine Ecology Progress Series}, 487:185--200, 2013.

\bibitem{PRAIRIE2015}
J.~C. Prairie, K.~Ziervogel, R.~Camassa, R.~M. McLaughlin, B.~L. White, C.~Dewald, and C.~Arnosti.
\newblock Delayed settling of marine snow: Effects of density gradient and particle properties and implications for carbon cycling.
\newblock {\em Marine Chemistry}, 175:28--38, 2015.

\bibitem{Prarie2017}
J.~C. Prairie, K.~Ziervogel, R.~Camassa, R.~M. McLaughlin, B.~L. White, Z.~I. Johnson, and C.~Arnosti.
\newblock Ephemeral aggregate layers in the water column leave lasting footprints in the carbon cycle.
\newblock {\em Limnology and Oceanography Letters}, 2(6):202--209, 2017.

\bibitem{semcesen2021}
P.~O. Semcesen and M.~G. Wells.
\newblock Biofilm growth on buoyant microplastics leads to changes in settling rates: Implications for microplastic retention in the great lakes.
\newblock {\em Marine Pollution Bulletin}, 170:112573, 2021.

\bibitem{smith1992}
D.~C. Smith, M.~Simon, A.~L. Alldredge, and F.~Azam.
\newblock Intense hydrolytic enzyme activity on marine aggregates and implications for rapid particle dissolution.
\newblock {\em Nature}, 359(6391):139--142, 1992.

\bibitem{stokes1851}
G.~G. {Stokes}.
\newblock {On the Effect of the Internal Friction of Fluids on the Motion of Pendulums}.
\newblock {\em Transactions of the Cambridge Philosophical Society}, 9:8, Jan. 1851.

\bibitem{sutherland2023}
B.~R. Sutherland, M.~DiBenedetto, A.~Kaminski, and T.~van~den Bremer.
\newblock Fluid dynamics challenges in predicting plastic pollution transport in the ocean: A perspective.
\newblock {\em Phys. Rev. Fluids}, 8:070701, Jul 2023.

\bibitem{tatsii2023}
D.~Tatsii, S.~Bucci, T.~Bhowmick, J.~Guettler, L.~Bakels, G.~Bagheri, and A.~Stohl.
\newblock Shape matters: Long-range transport of microplastic fibers in the atmosphere.
\newblock {\em Environmental Science \& Technology}, 58(1):671--682, 2024.

\bibitem{yick2009}
K.~Y. Yick, C.~R. Torres, T.~Peacock, and R.~Stocker.
\newblock Enhanced drag of a sphere settling in a stratified fluid at small reynolds numbers.
\newblock {\em Journal of Fluid Mechanics}, 632:49–68, 2009.

\bibitem{Zhang2019}
J.~Zhang, M.~J. Mercier, and J.~Magnaudet.
\newblock Core mechanisms of drag enhancement on bodies settling in a stratified fluid.
\newblock {\em Journal of Fluid Mechanics}, 875:622–656, 2019.

\bibitem{zvirin1975}
Y.~Zvirin and R.~Chadwick.
\newblock Settling of an axially symmetric body in a viscous stratified fluid.
\newblock {\em International Journal of Multiphase Flow}, 1(6):743--752, 1975.

\end{thebibliography}

\end{document}